\newcommand{\thp}{$\Theta^+$}

\documentclass[
    ,final            
  ]
  {aipproc}

\layoutstyle{6x9}

\begin{document}

\title{Review of Recent Jlab Results}


\author{Elton S. Smith}{
  address={Thomas Jefferson National Accelerator Facility, Newport News, Virginia 23606, USA}
}

\begin{abstract}
High quality polarized electron beams at Jefferson Lab make possible
precision measurements of hadronic properties in the regime of strongly
interacting QCD.  We will describe a few programs at Jefferson Lab that
are making measurements that link the basic static properties of hadrons
to their quark sub-structure. For example, parity-violating electron proton
elastic scattering probes the spatial distribution of strange quarks in
the nucleon. The nucleon-Delta transition form factors give us information
about the deformation of nucleons and Deltas.  Finally, new high
statistics measurements of photons scattering off proton and deuteron
targets are used to set upper limits on the production of exotic
baryons with strangeness S=+1.  These examples will be used to illustrate
the capabilities and focus of the experimental program at JLab. 
\end{abstract}

\maketitle


\section{Introduction to JLab}

The Continuous Electron Beam Accelerator Facility (CEBAF) at the 
Thomas Jefferson National Accelerator Facility (Jefferson Lab) is devoted to the 
investigation of the electromagnetic structure of mesons, nucleons, 
and nuclei using high energy and high duty-cycle 
electron and photon beams. 

CEBAF is a superconducting electron accelerator \cite{CEBAF} 
with a maximum energy of 6~GeV and 100~\% duty-cycle.  Three 
electron beams with a maximum total current of 200~$\mu A$ can be used 
simultaneously for electron scattering experiments in the experimental 
areas. The accelerator design concept is based on two parallel 
superconducting continuous-wave linear accelerators joined by magnetic 
recirculation arcs. The accelerating structures are five-cell 
superconducting niobium cavities with a nominal average energy gain of
5~MeV/m. The accelerator operation has met all design goals, achieving
5.7 GeV for physics running, and delivering three high quality beams with intensity
ratios exceeding 10$^6$:1. The electron beam is
produced using a strained GaAs photocathode allowing the delivery of polarized electrons
(P$_{e}~\geq$ 75\%) simultaneously to all three halls.

Three experimental areas are available for simultaneous
experiments, the only restriction being that the beam energies have to be 
multiples of the single pass energy. The halls contain
complementary equipment which cover a wide range of physics topics: Hall A \cite{halla}
has two high resolution magnetic spectrometers with 10$^{-4}$ 
momentum resolution in a 10\% momentum bite, and a solid angle of 
8~msr. Hall B houses the large 
acceptance spectrometer, CLAS \cite{clas}. Hall C 
uses a combination of a high momentum
spectrometer ($10^{-3}$ momentum resolution, 7~msr solid angle and maximum
momentum of 7 GeV/c) and a short orbit spectrometer. 

To illustrate the physics which is being addressed at Jefferson Lab, we have chosen
three topics of current interest: measurements of parity violation 
in electron elastic scattering, the N-$\Delta$ transition form factors,
and high statistic searches for the $\Theta^+$ exotic strangeness +1 baryon with the CLAS.
These programs have the common goal of probing baryon structure which comes from 
quark pairs beyond the contribution from the standard three valence quarks. 

\section{Strange quarks in the nucleon}

Early deep-inelastic scattering experiments demonstrated that the structure
of the proton can not be described by its $uud$ valence structure alone. For example,
quarks carry only 50\% of the proton momentum and are responsible for less than 30\% its 
spin \cite{spinannrev}. The fraction and distribution of $s$ and $\overline{s}$ quarks 
were also determined in $\nu$ and $\overline{\nu}$ scattering experiments. For example, NuTeV \cite{NuTeV}
reports the fraction $(s+\overline{s})$/$(\overline{u}+\overline{d})$ to be 
0.42 $\pm$ 0.07 $\pm$ 0.06 at a $Q^2$ of 16 GeV$^2$.
But most of these sea quarks are very short lived ($\sim \hbar/\sqrt{Q^2}$) and arise
from perturbative evolution of QCD \cite{burkardt92}. 
The focus of this section is to describe experiments that are sensitive to
the long-lived component of $s\overline{s}$ pairs which 
contribute to the static properties of the nucleon such as its magnetic moment and charge radius.

It is easy to understand how $s\overline{s}$ pairs in the sea can contribute to the 
proton magnetic moment in a simple hadronic picture \cite{hannelius}. During the process of 
$uuds\overline{s}$ fluctuations, the quarks
will tend to arrange themselves into energetically favorable configurations, the
lowest state being a $\Lambda K^+$. The $K^+$ must be emitted in an L=1 state to
conserve angular momentum and parity. The probability that the $\Lambda$ has its spin
anti-parallel to the proton is twice as likely as the probability that the spin
have the same direction. In this case, the (positive) kaon will have $l_z=+1$
and contribute a positive amount to the magnetic moment of the proton. In addition,
the spin of the (negative) s quark in the $\Lambda$ is anti-aligned with the proton spin so
it also gives a positive contribution to the magnetic moment. By convention, however,
the usual definition of the strange magnetic moment ($G_M^s(Q^2=0)$) does not include the
s-quark charge of ($-\frac{1}{3}$), and therefore this picture actually predicts a \emph{negative}
strange magnetic moment. 
Over the past decade there
have been many different models \cite{pvmodels} used to estimate both the magnitude and sign of $G_M^s$,
but all have been forced to make non-trivial approximations. Most models predict a negative $G_M^s$,
as in the case of the naive hadronic model above, with a value in the range of $-0.8$~to~$0$. But
it is clear for the moment that experimental measurements are required to guide
our understanding of the contributions of strange quarks to the properties of the proton.

Parity violation in electron scattering is a unique tool for probing
the contribution of $s\overline{s}$ sea quarks to the structure of the nucleon.
The weak interaction is probed by measuring the electron helicity dependence of 
the elastic scattering rate off unpolarized targets.  
Experimental sensitivities of
the order of a part per million are required to measure the parity violating asymmetry
on a nucleon which can be expressed as
\begin{eqnarray}
A^p_{PV} & = & {{d\sigma_R - d\sigma_L} \over {d\sigma_R - d\sigma_L}} 
= {{G_F Q^2} \over {4\pi\alpha\sqrt{2}}} \left[ {A_E + A_M + A_A} \over 
\epsilon \left(G_M^\gamma\right)^2 + \tau\left(G_M^\gamma\right)^2 \right] , \label{eq:eq1}
\end{eqnarray}
where $A_E = \epsilon G_E^Z(Q^2) G_E^\gamma(Q^2)$, $A_M = \tau G_M^Z(Q^2) G_M^\gamma(Q^2)$ are
the electric and magnetic $\gamma$-Z interference terms, $\tau = Q^2/4M_N^2$, and $\epsilon$
is the polarization of the virtual photon. The isolation of the electric and magnetic
terms can be accomplished by taking advantage of the sensitivity to kinematics 
(though $\epsilon$ and $\tau$) and taking data 
at different scattering angles. The last term, $A_A$, picks out the
contribution due to the axial form factor, but the forward-angle measurements are insensitive to this
term, so we refer the reader to our references for details \cite{sample}. 
Finally, the flavor decomposition of the form factors
can be determined uniquely assuming charge symmetry and by combining parity-violating
asymmetries with measurements of the electric and magnetic form factors on the proton and neutron.
The scattering
amplitude from a zero spin isoscalar target, such as $^4He$, is particularly simple because
it does not allow magnetic transitions so the asymmetry is sensitive only to the isoscalar electric form factor:
\begin{eqnarray}
A^{He}_{PV} & = & {{G_F Q^2} \over {4\pi\alpha\sqrt{2}}} \left[ 4\sin^2{\theta_W} + 
{G_E^s \over {\frac{1}{2} \left( G_E^{\gamma p} + G_E^{\gamma n} \right)}} \right] , \label{eq:eq2}
\end{eqnarray}
The contribution of the asymmetry which is attributable to strange quarks is obtained by
subtracting the calculated asymmetry assuming no strange contribution, including
radiative corrections and best estimates of the electromagnetic form factors. 
The uncertainties in the calculations are estimated
and included separately when quoting the uncertainties in the strange form factors. 
Thus a fairly extensive experimental program is required to isolate all contributions.

\begin{figure}
   \includegraphics[width=10cm]{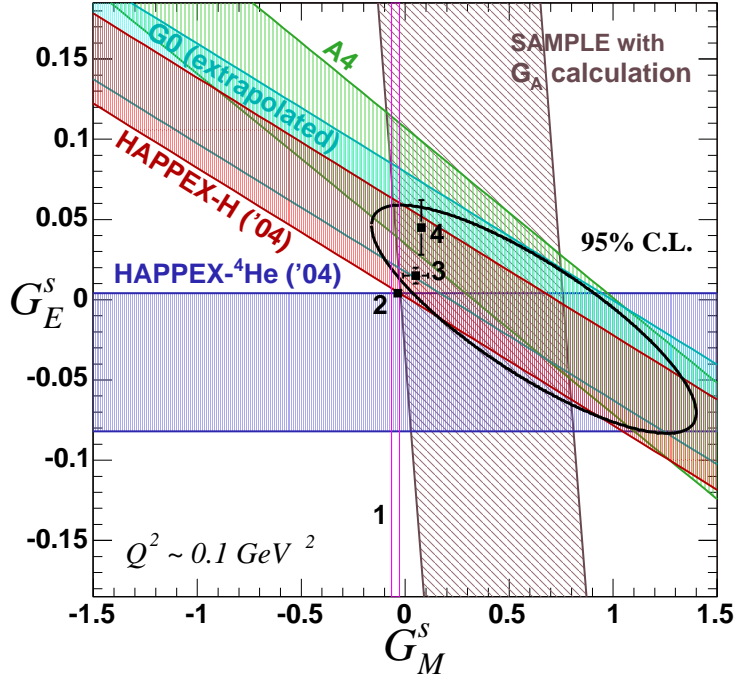}
  \caption{Constraints on the $G_E^s$ and $G_M^s$ strange form factors at $Q^2 = 0.1GeV^2$. The 
curves give the 95\% contours. 
Sample model calculations are also indicated 
as 1 \cite{pvtheory1} 2 \cite{pvtheory2} 3 \cite{pvtheory3} and 4 \cite{pvtheory4}. \label{fig:pv1}}
\end{figure}   

There are two major experimental programs at JLab studying parity violation:
HAPPEX \cite{happexprc} and G0 \cite{G005}. Both collaborations have 
recently reported results. The precision of
these experiments, together with other experimental and theoretical efforts world-wide 
\cite{sample,A404,A405,parityannrev,parityprog,emworkshop}, has now achieved sufficient 
accuracy to be sensitive to the contribution
of strange quarks to the nucleon form factors.  
The first measurements of parity-violating electron scattering at JLab were carried out by the HAPPEX
collaboration. The experiment detected the scattered electrons in each of the
Hall A high resolution spectrometers, with sufficient hardware resolution to spatially
separate the elastically scattered events from inelastic events into special
lead-scintillator absorption counters whose output was integrated during a 30 ms window.
Window pairs of opposite helicity were 
chosen randomly at 15 Hz. A half wave plate in the CEBAF injector was inserted
or removed about once a day, switching the helicity of the beam without changing the electronics.
HAPPEX has previously reported measurements
of parity violation in ep elastic scattering at Q$^2$=0.48 GeV$^2$ \cite{happex01}.
The asymmetry errors are completely dominated by statistical uncertainties, and the helicity
correlated systematic uncertainties are about 0.1 ppm. 
Recently it has completed asymmetry measurements at Q$^2$=0.1 GeV$^2$ in ep scattering 
giving $G_E^s + 0.080\;G_M^s = 0.030\pm0.025\pm0.006\pm0.012$ \cite{happex05} and
a measurement of elastic scattering on $^4$He at a Q$^2$=0.091 GeV$^2$ which yields a direct
measurement of $G_E^s = -0.038\pm0.042\pm0.010$, consistent with zero.
\cite{happexhe4}. These 
measurements were performed at electron scattering angles of 6$^\circ$, made possible by the new
superconducting septum magnets installed in Hall A \cite{septum}. We note that HAPPEX
continues to take data and is expected to improve the statistical uncertainty in each measurement
by a factor of 2.5--3 this year.
 
The G0 experiment uses a dedicated detector \cite{G0detector} to determine $G_E^s$, $G_M^s$, and 
$G_A^e$ over a broad kinematical range. To date measurements
have been completed for forward angles. The result of their first experimental run has been 
the determination of the linear combination $G_E^s + \eta G_M^s$ for $Q^2$ between 0.12 to 1.0 GeV$^2$
($\eta = \tau G_M^p/\epsilon G_E^p$).
These measurements are consistent with HAPPEX and A4,
but they provide broad coverage in $Q^2$ to probe the spatial dependence of
strange quarks in the nucleon. The three lowest $Q^2$ points (out of 16 measurements)
can be used together with the other experiments (see  Fig.\,\ref{fig:pv1}) to 
determines a central value for $G_E^s$ and $G_M^s$ separately for $Q^2=0.1$ GeV$^2$.

We close by summarizing the experimental situation, and relating the measured quantities back 
to our simple hadronic picture for possible strange contributions to the proton magnetic moment.
Analysis of the world measurements leads to $G_M^s = 0.55\pm0.28$ and
$G_E^s = -0.01\pm0.03$. To compare to the proton magnetic moment of +2.79 n.m.
we multiply $G_M^s$ by ($-\frac{1}{3}$) and, taken at face value,
it is approximately 7\% of the magnitude of the proton magnetic moment,
and opposite in sign. The interpretation of positive values for $G_M^s$ has been investigated
within the context of the quark model \cite{zou}. This study shows that, even though the quark
model is very successful at predicting relations between baryon magnetic moments, the quark model
does not lead naturally to positive values for $G_M^s$. The authors find that positive moments are generated when
the $\overline{s}$ is in the ground state and the $uuds$ quarks are in excited states. Although
these configurations do not seem very natural, they are very analogous to quark-models configurations
for pentaquarks, which is the topic of the last section of this paper.

\section{N-$\Delta$ deformation}

Within the spin-flavor SU(6) symmetry of the quark model, the $\Delta$ resonance is 
a completely symmetric object with all quark spins aligned. Within this model, the
photo-excitation $\gamma^* N \rightarrow \Delta(1232) \rightarrow N\pi$ proceeds via
a single quark spin flip in the L=0 nucleon ground state. This magnetic dipole transition,
characterized by the M$_{1+}$ multipole, predicts negligible contributions from the
electric (E$_{1+}$) and scalar (S$_{1+}$) quadrupole multiples. However, strong-interaction
dynamics will modify this picture, and several models predict deviations from this simple
picture which would indicate deformations of both the nucleon and the $\Delta$ \cite{henley01}. 
For example, the deformation
of the nucleon and $\Delta$ could result from D-state admixtures into the baryon wave function.
However, D-wave admixtures are expected to be small, but quark-antiquark pairs which are present in the 
nucleon can also contribute to the quadrupole moment. These latter contribute to the 
quadrupole moment via a two-body spin tensor in the charge operator, even when the valence quarks 
are in pure S states. The physical interpretation in this case is that the distribution of
$q \overline{q}$ sea quarks in the nucleon deviate from spherical symmetry. These models interpret
the negative values of the measured transition quadrupole moments as corresponding to
a positive intrinsic quadrupole moment (cigar shape) for the nucleon and a negative moment 
(pancake shape) for the $\Delta$.

Complete angular distributions of the pion in the reaction $e p \rightarrow ep \pi^0$ at
the peak of the $\Delta$ allow the experimental determination of the quadrupole moments.
The differential cross section of a polarized beam (helicity $h=\pm 1$) and unpolarized target 
can be written as a function
of the transverse and longitudinal structure functions $\sigma_T$ and  $\sigma_{L}$, 
and the interference terms $\sigma_{TT}$, $\sigma_{LT}$ and $\sigma_{LT'}$ :
\begin{eqnarray}
{d^2\sigma \over d \Omega_\pi^*} & = & {p_\pi^* \over k_\gamma^*}
\left( \sigma_T + \epsilon_L\sigma_L + \epsilon\sigma_{TT} \; \sin^2{\theta_\pi^*} \cos{2\phi_\pi^*} \right. \\
 & & + \left. \sqrt{2\epsilon_L\left(\epsilon + 1 \right)} \;\sigma_{LT}\;\sin{\theta_\pi^*}\cos{\phi_\pi^*} 
 +{} h\;\sqrt{2\epsilon_L\left(\epsilon - 1 \right)} \;\sigma_{LT'}\;\sin{\theta_\pi^*}\sin{\phi_\pi^*}
\right), \;\nonumber
\end{eqnarray}
where $p_\pi^*$, $\theta_\pi^*$, and $\phi_\pi^*$ are the pion center-of-mass (c.m.) momentum and 
angles, $\epsilon$ and $\epsilon_L$
are the usual transverse and longitudinal polarizations of the virtual photon, and $k_\gamma^*$ is the
real-photon-equivalent c.m. energy. 
The moments of interest can be selected as terms
in the partial wave expansion of the structure functions using 
Legendre polynomials \cite{burkertlee,burkert05}. Unique solutions can be obtained by truncating the 
multipole expansion to terms involving only $M_{1+}$.
Fitting for the coefficients of the expansion allows the extraction 
of the electric ($E_{1+}$) and scalar ($S_{1+}$) quadrupole moments through their interference
with the dominant $M_{1+}$. The systematic errors due to the truncation of higher multipoles are estimated
by calculating the effects of higher partial waves using realistic parameterizations
of higher mass resonances and backgrounds. At the peak of the $\Delta$ and for $Q^2$ of a few
GeV$^2$, this procedure results in a largely model-independent extraction of the moments.
 
Early interest in the quadrupole moments stemmed from the perturbative QCD predictions of
quark helicity conservation which requires that $E_{1+} \rightarrow M_{1+}$ 
and $S_{1+}$ approaches a constant at very high $Q^2$.
Therefore, the first experiments at JLab in Hall C \cite{frolov99} determined the quadrupole moments
at the highest accessible $Q^2 \sim 3-4$ GeV$^2$, followed by 
results from CLAS \cite{joo02} at intermediate momentum transfer. These measurements showed that 
$E_{1+}$ was not only a few percent of $M_{1+}$, but also of opposite sign. The asymptotic 
kinematics of perturbative QCD was clearly well beyond the range of these measurements.

\begin{figure}
  \includegraphics[width=10cm]{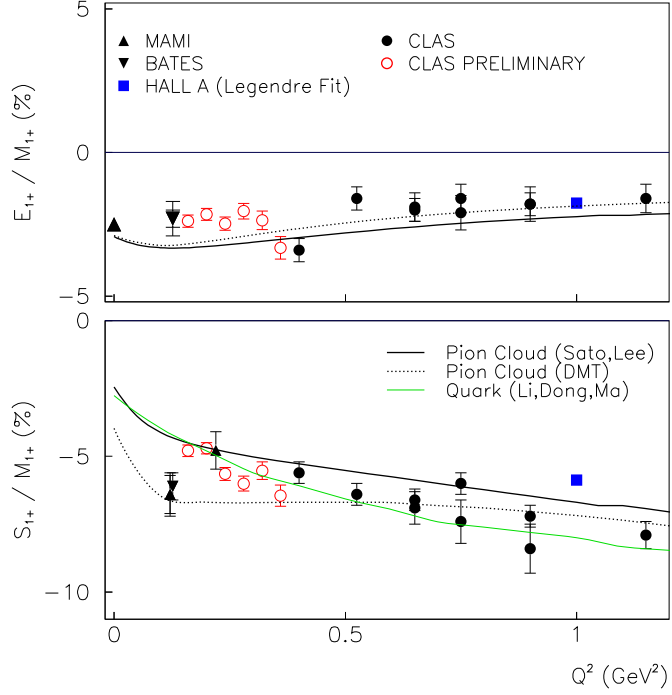}
  \caption{$Q^2$ dependence of the ratios of the electric (E$_{1+}$) and scalar (S$_{1+}$) quadrupoles
to the dominant dipole moment. The curves are model calculations from references 
\cite{satolee}, \cite{dmtmodel}  and \cite{lidongma}. \label{fig:remrsm}}  
\end{figure}   

Recently the focus has shifted to understanding the deformation of the nucleon at 
very low $Q^2$ which probes chiral symmetry breaking, models of heavy baryon chiral
perturbation theory, and the shape of the pion cloud surrounding the nucleon. Data at $Q^2 = 0.12$~GeV$^2$ from
MAMI \cite{mami01} and Bates \cite{bates01} show a surprisingly large scalar quadrupole moment of 6.5\%,
twice that expected from the trend of higher energy measurements and most models.
New data from CLAS \cite{lcsmith05}
were taken with an electron beam energy of 1.046 GeV, polarization of $\sim$70\% 
and 10 nA beam current on a 2-cm liquid hydrogen target. The $\pi^0$ electroproduction data from
this experiment have provided nearly complete angular distributions for $Q^2 = 0.14-0.38$ GeV$^2$ 
and $W = 1.2-1.3$ GeV. The broad acceptance of the CLAS detector allows measurements of
the full angular distribution over a broad range of $Q^2$ and covering the W range of the $\Delta$.
The ratios of  $E_{1+}/M_{1+}$ and $S_{1+}/M_{1+}$ from this experiment along with other
measurements are shown in Fig.\,\ref{fig:remrsm}. Important constraints on the non-resonant
backgrounds have been obtained by measuring the helicity-dependent structure function $\sigma_{LT'}$ 
\cite{joo03}. Weak non-resonant backgrounds underlying the $\Delta$ peak can be enhanced through their interference
with the magnetic multipole $Im{M_{1+}}$, which is supressed in $\sigma_{LT}$ due to the
vanishing of the real part of the resonant amplitude at the pole.

In summary, there is general agreement on the value of $E_{1+}/M_{1+}\sim$ -2.5\%, which is approximately constant at
small $Q^2$. The recent analysis of CLAS indicates that $S_{1+}/M_{1+}$ ratio shows a 
$Q^2$-dependence in this same kinematic region, in apparent disagreement with previous measurements that
show it is leveling off. Global parameterizations of electroproduction data is underway and will
be used along with $\pi^+$ data to extract the nucleon parameters with refined accuracy.


\section{Search for pentaquarks}

Pentaquark states have been studied both theoretically and 
experimentally for many years \cite{pdg86}. Recent interest was revived by
predictions within the chiral soliton model \cite{diakonov} for the
existence of an anti-decuplet of 5-quark resonances with spin and parity $J^\pi = \frac{1}{2}^+$. 
The lowest mass member of the anti-decuplet, now called the \thp, is an isosinglet with valence quark 
configuration $uudd\bar{s}$ with strangeness $S=+1$, was predicted to have a mass of 1.53 GeV/c$^2$ and a 
width of $\sim 0.015$ GeV/c$^2$. These definite predictions
prompted experimental searches to focus attention in this 
mass region. Shortly after the first observation of an exotic $S=+1$ state
by the LEPS collaboration \cite{leps03}, many experiments 
presented confirming evidence of the initial report, including
two observations using the CLAS detector \cite{clas-d,clas-p}. However, there have
been an increasing number of experiments reporting null results,
even though in very different reactions and disparate kinematic regions \cite{dzierba04}.

This paper reports on two searches for a narrow $S=+1$ baryon in the mass range 
between 1520 and 1600 MeV using the CLAS detector.
These new high-statistics searches were conducted using the tagged photon beam facility in
Hall B \cite{tagger} with deuterium \cite{g10} and proton \cite{g11} targets.
The deuterium data can be compared directly with the first reported observation
by CLAS \cite{clas-d}, which analyzed a data set referred to as ``G2a.'' 
The G2a data set consisted of two run periods at E$_e$ = 2.478 GeV and 3.115 GeV,
corresponding to 0.83 and 0.35 pb$^{-1}$ respectively. The photon energy was tagged beginning below the
reaction threshold of 1.51 GeV up to 95\% of the electron beam energy. The target was 10 cm long and located
on CLAS center, and the CLAS torus field was set at 90\% of its maximum value.


\begin{figure}
\begin{minipage}[t]{7cm}
\begin{flushleft}
  \includegraphics[width=8cm]{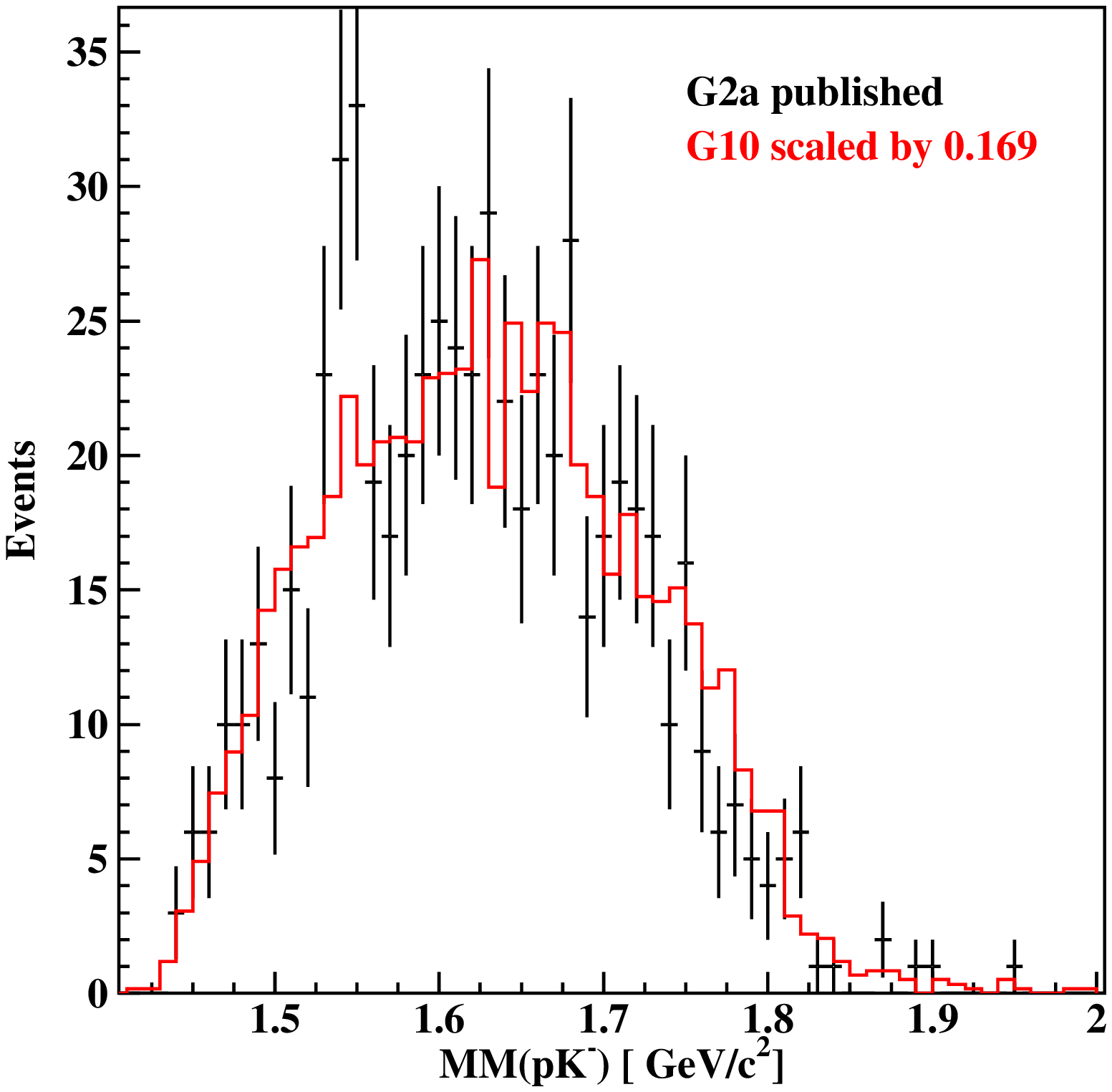}
\end{flushleft}
\end{minipage}
\hfill
\vspace{-7.5cm}
 
\begin{minipage}[t]{7cm}
\begin{flushright}
  \includegraphics[width=8cm]{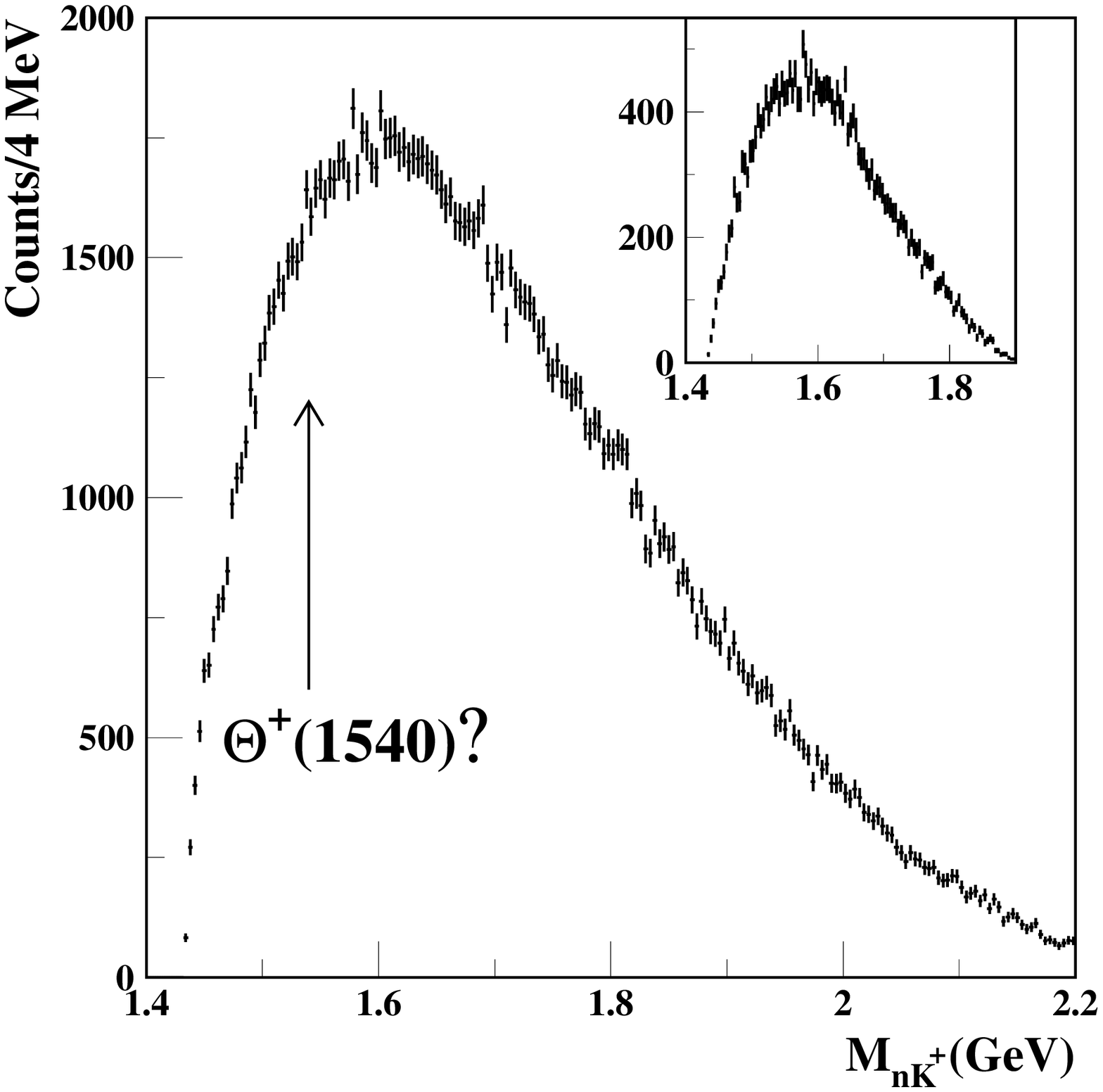}
\end{flushright}
\end{minipage}

  \caption{(Left) $MM(pK^-)$ distribution of the G10 data (red histogram) compared
           to the G2a distribution (black points with error bars). Selection
           cuts have been made on the G10 data set to best represent the
           experimental conditions of G2a and scaled for direct comparison. \label{fig:gdata}
           (Right)Distribution of the $nK^+$ mass spectrum for the G11 data.
           No evidence for narrow structures is apparent. The inset
           shows the mass distribution with selection cuts to reproduce the
           SAPHIR analysis \cite{saphir}. Note: All results from CLAS are preliminary.}
\end{figure}

The new deuterium data set, referred to as G10, used a primary electron beam energy of E$_e$ = 3.767 GeV creating
a tagged photon beam with energies E$_\gamma$ between 1 and 3.6 GeV. The run was
divided evenly between two field settings of the torus magnet (60\% and 90\% of the full-field
current of 3860 A). The hardware trigger required two charged particles in
any two sectors and the target was a 24-cm long cell contaning liquid deuterium 
and located 25 cm upstream of nominal CLAS center.
The integrated luminosity for this data set was 38 pb$^{-1}$ for E$_\gamma$ greater than 
1.5 GeV. Evidence for the exotic
baryon was searched for using the reaction $\gamma d \rightarrow K^- K^+ p n$.
The momentum of the charged particles was determined using magnetic analysis and
their mass determined using time-of-flight techniques. The analysis selected 
a detected proton, $K^+$ and $K^-$ in the final state, all originating from the same beam 
bucket. The neutron momentum and energy was reconstructed from the measured
charge particle tracks and known energy of the incident photon. The exotic 
$\Theta^+$ baryon was searched for in the decay $\Theta^+ \rightarrow K^+ n$, where
the $K^+$ uniquely identified the positive strangeness of the baryonic state.
The mass of the $K^+ n$ system is shown in Fig.\,\ref{fig:gdata}(left) for both the G2
and G10 data samples.
The spectrum is relatively smooth and does not exhibit the peak found in the original
smaller data set. The figure overlays the two spectra with selection cuts on G10 to
mimic the G2a data set as closely as possible including the photon energy spectrum.
If we use the G10 data as a background shape, the peak at 1.54 GeV in the G2 data set
is consistent with a 3 $\sigma$ fluctuation. The new data show no indication of
a peak in the $\Theta^+ \rightarrow K^+ n$ exotic S=+1 channel.

The measurement on the proton (G11) was conducted at lower photon energy and using a 
different reaction from the published CLAS result \cite{clas-p}. 
This new experiment used a primary electron beam energy E$_e$ = 4.0 GeV
to produce a tagged photon beam in the range of E$_\gamma$ between 1.6 and 3.8 GeV. 
The target consisted of a 
40-cm long cylindrical cell with liquid hydrogen. The data were taken over
a period of 50 days, corresponding to an integrated luminosity of 70 pb$^{-1}$,
which is an order-of-magnitude greater than previous experiments with tagged
photons. Evidence for the $S=+1$ exotic baryon $\Theta^+$ was searched for in
the reaction $\gamma p \rightarrow \overline{K}^0 K^+ n$. The search was
conducted for the baryon produced in association with the $\overline{K}^0$
and subsequent decay $\theta^+ \rightarrow K^+ n$. The positive strangeness of the baryon
was uniquely tagged using the $K^+$ of the decay. The charged kaon, as well as the pions
from the $\overline{K}^0 \rightarrow \pi^+\pi^-$ decay, were reconstructed in the 
CLAS detector. The neutron momentum was reconstructed using energy
and momentum conservation and the known energy of the incident photon. The missing mass peak
of the neutron was reconstructed to be 939 MeV with a width of 10 MeV.
The $\Sigma^+\rightarrow \pi^+n$ and $\Sigma^-\rightarrow \pi^-n$ reactions 
are also present in this data sample and are 
reconstructed with an accuracy of less than 1 MeV and resolution of 3 MeV. These
baryons, as well as the $\Lambda^*(1520)$ were excluded from the present analysis.
After all cuts, the reconstructed $n K^+$ mass spectrum is shown in Fig.\,\ref{fig:gdata}(right).
The spectrum is smooth and structureless, with no indication of a peak near 1540 MeV
where the $\Theta^+$ was previously reported \cite{saphir,ostrick}. An upper limit to the $\Theta^+$
cross section of less than 1 nb was set based on assuming a t-exchange production mechanism 
for the reaction. 

We conclude with a brief summary of the status of pentaquark measurements with CLAS.
There are two published observations \cite{clas-d,clas-p} of the $\Theta^+$ baryon in photon beam
reactions on deteurium and proton targets. The measurement on deuterium has been 
repeated with higher statistics and the new data do not confirm the original observation.
A new experiment to verify the production off the proton in the reaction 
$\gamma p \rightarrow \pi^+ K^- K^+ n$ for photon energies between 3 and 5.5 GeV
has been approved \cite{superg6}, but has not yet received beam time. However, a recent high statistics 
experiment on the proton at lower energy  \cite{g11pub} has searched for the narrow exotic baryon in the reaction 
$\gamma p \rightarrow \overline{K}^0 K^+ n$ and found no evidence for the state
in the $n K^+$ mass range between 1520 and 1600 MeV.





\begin{theacknowledgments}
I would like to thank Alberto dos Reis and all the organizers of the
conference for their gracious hospitality. I would also like
to thank K. Paschke, D. Armstrong and D. Beck of the HAPPEX and G0 collaborations
for making available the results 
on parity violation. From the CLAS collaboration I would like to
express special thanks to 
K. Joo and L. Smith for help in preparation
of materials on Delta production, and S. Stepanyan, R. De Vita and
M. Battaglieri for discussions of the pentaquark search.
The Southeastern Universities Research Association 
(SURA) operates the Thomas Jefferson National Accelerator Facility 
for the U.S. Department of Energy under contract DE-AC05-84ER40150.
\end{theacknowledgments}



\bibliographystyle{aipproc}   

\bibliography{hadron05}

\IfFileExists{\jobname.bbl}{}
 {\typeout{}
  \typeout{******************************************}
  \typeout{** Please run "bibtex \jobname" to optain}
  \typeout{** the bibliography and then re-run LaTeX}
  \typeout{** twice to fix the references!}
  \typeout{******************************************}
  \typeout{}
 }

\end{document}